\documentclass[12pt]{article}
\input{psfig}
\usepackage{xspace,amsmath,amssymb}
\usepackage{color}
\usepackage{mathrsfs}
\usepackage[colorlinks=true,linkcolor=blue,filecolor=blue,citecolor=blue,urlcolor=blue]{hyperref}
\newcommand{\namedref}[2]{\hyperref[#2]{#1~\ref*{#2}}}
\renewcommand{\eqref}[1]{\hyperref[#1]{~(\ref*{#1})}}

\newtheorem{thm}{Theorem}[section]      %
\newtheorem{lemma}[thm]{Lemma}

\newtheorem{corollary}[thm]{Corollary}
\newtheorem{proposition}[thm]{Proposition}

\newcommand{\commentout}[1]{}
\newcommand{\IN}{\mbox{$I\!\!N$}}
\newcommand{\IR}{\mbox{$I\!\!R$}}

\newcounter{deff}
\newcommand{\hide}[1]{}

\def\qed{\vrule height7pt width4pt depth1pt\par}

\newenvironment{gproof}{\noindent{\bf Proof Sketch:~~}}{\qed}
\newcommand{\BPF}{\begin{gproof}} \newcommand {\EPF}{\end{gproof}}
\newenvironment{fproof}{\noindent{\bf Proof:~~}}{\qed}
\newcommand{\BPRF}{\begin{fproof}} \newcommand {\EPRF}{\end{fproof}}
\newenvironment{hproof}{\noindent{\bf Proof~}}{\qed}
\newcommand{\BPR}{\begin{hproof}} \newcommand {\EPR}{\end{hproof}}

\newcommand{\BI}{\begin{itemize}}
\newcommand{\EI}{\end{itemize}}
\newcommand{\BE}{\begin{enumerate}}
\newcommand{\EE}{\end{enumerate}}

\newcommand{\BT}{\begin{thm}}   \newcommand{\ET}{\end{thm}}
\newtheorem{dfn}[thm]{Definition}      %
\newcommand{\BD}{\begin{dfn}}   
\newcommand{\ED}{\end{dfn}}
\newtheorem{corr}[thm]{Corollary}      %
\newcommand{\BCR}{\begin{corr}} \newcommand{\ECR}{\end{corr}}
\newtheorem{constr}[thm]{Construction}
\newcommand{\BCT}{\begin{constr}} \newcommand{\ECT}{\end{constr}}
\newtheorem{prop}[thm]{Proposition}
\newcommand{\BP}{\begin{prop}}   \newcommand{\EP}{\end{prop}}
\newtheorem{lemm}[thm]{Lemma}   %
\newcommand{\BL}{\begin{lemm}}   \newcommand{\EL}{\end{lemm}}
\newtheorem{clm}[thm]{Claim}            %
\newcommand{\BCM}{\begin{clm}}   \newcommand{\ECM}{\end{clm}}
\newtheorem{sclm}[thm]{Sub-Claim}            %
\newcommand{\BSCM}{\begin{sclm}}   \newcommand{\ESCM}{\end{sclm}}
\newtheorem{assumption}[thm]{Assumption}            %
\newcommand{\BA}{\begin{assumption}}   \newcommand{\EA}{\end{assumption}}
\newtheorem{remark}[thm]{Remark}            %
\newcommand{\BR}{\begin{remark}}   \newcommand{\ER}{\end{ER}}
\newtheorem{example}[thm]{Example}

\newcommand{\complex}{\canon}
\newcommand{\complexity}{\complex}

\newcommand{\M}{{\cal M}}

\newcommand{\T}{{T}}

\newcommand{\Z}{{\cal Z}}

\newcommand{\bitset}{\{0,1\}}

\newcommand{\Qchange}{q}

\DeclareRobustCommand*{\slashfracstyle}[1]{%
  {\ensuremath{\mbox{\fontsize\sf@size\z@\selectfont #1}}}}

\DeclareRobustCommand*{\slashfrac}[2]{\leavevmode
  \raise.5ex\hbox{\scriptsize #1}\kern-.13em/%
  \kern-.15em\lower.25ex\hbox{\scriptsize #2}}

\def\*{\Z^*}

\def\B{\{0,1\}}
\def\B*{\B^*}

\def\union{\cup}

\newcommand{\canon}{\cC}

\def\cC{\mathscr{C}}

\def\cH{{\cal H}}

\def\cN{{\cal N}}

\def\cQ{{Q}}

\newcommand{\fullv}[1]{#1}
\newcommand{\shortv}[1]{\commentout{#1}}

\newcommand{\beginsmall}[1]{\vspace{-.1em}\begin{#1}\itemsep-\parskip}

\newenvironment{RETHM}[2]{\trivlist \item[\hskip 10pt\hskip\labelsep{\bf
#1\hskip 5pt\relax\ref{#2}.}]\it}{\endtrivlist}
\newcommand{\rethm}[1]{\begin{RETHM}{Theorem}{#1}}
\newcommand{\repro}[1]{\begin{RETHM}{Proposition}{#1}}
\newcommand{\relem}[1]{\begin{RETHM}{Lemma}{#1}}
\newcommand{\recor}[1]{\begin{RETHM}{Corollary}{#1}}
\newcommand{\erethm}{\end{RETHM}}
\newcommand{\erepro}{\end{RETHM}}
\newcommand{\erelem}{\end{RETHM}}
\newcommand{\erecor}{\end{RETHM}}

\newcommand{\bbox}{\vrule height7pt width4pt depth1pt}

\fullv{
\usepackage{chicagor}
\bibliographystyle{chicagor}}
\shortv{\usepackage{named}
\bibliographystyle{named}}

\newcommand{\bott}{\omega}
\begin{document}
\title{Sequential Equilibrium in Computational Games}
\author{
Joseph Y. Halpern\thanks{Supported in part
by NSF grants 
IIS-0812045, IIS-0911036, 
and CCF-1214844, AFOSR	grants FA9550-08-1-0266 and
FA9550-12-1-0040, and 
ARO	grant W911NF-09-1-0281.}\\
Cornell University\\
halpern@cs.cornell.edu 
\and Rafael Pass\thanks{
Supported in part by an Alfred P. Sloan Fellowship, a
Microsoft Research Faculty Fellowship, 
NSF Awards CNS-1217821 and CCF-1214844, NSF CAREER Award CCF-0746990,
AFOSR YIP Award FA9550-10-1-0093, and DARPA and AFRL under contract
FA8750-11-2-0211. The views and conclusions contained in this document
are those of the authors 
and should not be interpreted as representing the official policies,
either expressed or implied, of the Defense Advanced Research Projects
Agency or the US Government. 
}\\
Cornell University\\
rafael@cs.cornell.edu}
\maketitle

\begin{abstract}
We examine sequential equilibrium in the context of \emph{computational games}
\cite{HP08}, where agents are charged for computation.  In such games,
an agent can rationally choose to forget, so issues of imperfect recall
arise.  In this setting, we consider two notions of 
sequential
equilibrium.  One is an \emph{ex ante} notion, where a player chooses his
strategy before the game starts and is committed to it, but chooses
it in such a way that it remains optimal even off the equilibrium path.
The second is an \emph{interim} notion, where a player can reconsider at
each information set whether he is doing the ``right'' thing, and 
if not, can change his strategy.
The two notions agree in games of perfect recall, but not in games of
imperfect recall.   Although the interim notion seems more appealing,
\fullv{Halpern and Pass \citeyear{HP09} argue}
\shortv{in \cite{HP09} it is argued}
that there are some deep
conceptual problems with it in standard games of imperfect recall.  We
 show that the conceptual problems largely disappear in the
computational setting.  Moreover, in this setting, 
under natural assumptions, the two notions coincide.
\end{abstract}

\section{Introduction}
In \cite{HP08}, we introduced a framework to capture the idea
that doing 
costly computation affects an agent's utility in a game.  The approach,
a generalization of an approach taken 
\fullv{by Rubinstein \citeyear{Rub85},}
\shortv{in \cite{Rub85}},
assumes that players choose a Turing machine (TM) to play for them.
We consider Bayesian games, where each
player has a \emph{type}) (i.e., 
some private information); a player's
type is viewed as the input to his TM.  Associated with each TM $M$ and
input (type) $t$ is its \emph{complexity}.  The complexity 
could represent the running time of or space used by $M$ on input $t$.
While this is perhaps the most natural interpretation of complexity, it
could have other interpretations as well.  For example, it can be used
to capture the complexity of $M$ itself (e.g., the number of
states in $M$, which is essentially the complexity measure considered by
Rubinstein, 
who assumed that players choose a finite automaton to play for them
rather than a TM) or to model the cost of searching for a better
strategy (so that there is no cost for using a particular TM $M$,
intuitively, the strategy that the player has been using for years, but
there is a cost to switching to a different TM $M'$).  A player's
utility depends both on the actions chosen by all the players' machines
and the complexity of these machines.

This framework allows us to \fullv{, for example,} consider the tradeoff in a
game like \emph{Jeopardy} between choosing a strategy that spends longer
thinking before pressing the buzzer and one that answers quickly but is
more likely to be incorrect.  Note that if we take ``complexity'' here
to be running time, an agent's utility depends not only 
on
the complexity
of the TM that he chooses, but also on the complexity of the TMs chosen
by other players.  
We defined a
straightforward extension of Bayesian-Nash equilibrium in such
machine games, and showed that it captured a number of
phenomena of interest. 

Although in Bayesian games players make only one move,
a player's TM is doing some computation during the game.  This means
that solution concepts more traditionally associated with extensive-form
games, specifically, \emph{sequential equilibrium} \cite{KW82}, also turn
out to be of interest, since we can ask whether an agent wants to switch
to a different TM during the computation of the TM that he has chosen
(even at points off the equilibrium path).  
We can certainly imagine that, at the beginning of the computation, an
agent may have decided to invest in doing a lot of computation, but
part-way through the computation, he may have already learned enough to
realize that further computation is unnecessary.  In a sequential
equilibrium, intuitively, the TM he chose should already reflect this. 
It turns out that, even
in this relatively simple setting, there are a number of subtleties.  

The ``moves'' of the game that we consider are the 
outputs of the TM.
But what are the information sets?  We take them to be
determined by the states of the TM.  While this is a natural
interpretation, since we can view the TM's state as characterizing the
knowledge of the TM, it means that the information sets of the game are
not given exogenously, as is standard in game theory; rather, they are
determined endogenously by the TM chosen by the agent.%
\footnote{We could instead consider a ``supergame'', where at
the first step the agent chooses a TM, and then the TM plays for the
agent.  In this supergame, the information can be viewed as exogenous,
but this seems to us a less natural model.}
Moreover, in general, the game is one
of imperfect recall. An agent can quite rationally choose to forget (by
choosing a TM with fewer states, that is thus not encoding the whole
history) if there is a cost to remembering.   

Thinking of players as TMs can help clarify some issues when
considering games of imperfect information.  Consider the following
game, introduced by Piccione and Rubinstein \citeyear{PR97}:

\begin{figure}[htb]
\centerline{\psfig{figure=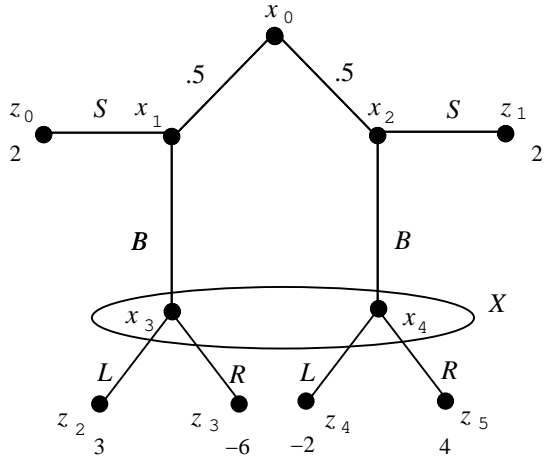,height=2.5in}}
\caption{A game of imperfect recall.}
\label{fig2}
\end{figure}

It is not hard to show that the strategy that maximizes expected utility
chooses action $S$ at node $x_1$, action $B$ at node
$x_2$, and action $R$ at the information set $X$ consisting of $x_3$ and
$x_4$.  Call this strategy $f$.  Suppose that the agent uses strategy
$f$.  If the agent knows his strategy (the typical assumption in game
theory), then the agent knows, when he is at the information set $X$,
then he must be at $x_4$.  So in what sense are $x_3$ and $x_4$ in the
same information set?  

Halpern \citeyear{Hal15} already observes that, in order to analyze
games of imperfect recall, we must make explicit what an agent knows
(including things like whether he knows his strategy, and whether he
recalls that he has switched strategies).  These issues are made
explicit when we consider TMs, and take an agent's information set to be
determined by the state of his TM---the state explicitly determines
what the agent knows (and remembers). Furthemore, although the
information sets are determined by the TM chosen by the agent, we can
force the agent into a situation of imperfect recall by ``charging'' a
lot for memory. Thus, our computational model generalizes standard
models of imperfect recall (and also perfect recall) while providing
an, in our eyes, more natural and explicit formalization of the game.

Games like that in Figure~\ref{fig2} are just an example of the
subtleties that must be dealt with when defining sequential equilibrium in
We
give a definition of sequential equilibrium 
in a companion paper \cite{HP09} for standard games of imperfect recall
that we extend here to take computation into account.  We show that, in
general, 
sequential equilibrium does not exist, but give simple conditions that 
guarantee that it exists, as long as a NE (Nash equilibrium) exists.
(As is shown 
by a simple example in \cite{HP08}, reviewed below, NE is not guaranteed to
exist in machine games, although sufficient conditions are given
to guarantee existence.)  

The definition of sequential equilibrium in \cite{HP09}
views sequential equilibrium as an \emph{ex ante} notion.   The 
idea is that a player chooses his strategy before the game
starts and is 
committed to it, but he chooses it in such a way that it remains
optimal even off the equilibrium path.  This, unfortunately, does
\emph{not} correspond to the more standard intuitions behind sequential
equilibrium, where players are reconsidering at each information set
whether they are doing the ``right'' thing, and 
if not, can change their strategies.  This
\emph{interim} notion of 
sequential rationality agrees with the ex ante notion in games of
perfect recall, but the two notions differ in games of imperfect
recall.  We argue in \cite{HP09}
there are some deep conceptual problems with
the interim notion
in standard games of imperfect recall.
We consider both an
ex ante and interim notion of sequential equilibrium
here.  We show that the conceptual problems when the game tree is given
(as it is in standard games) largely disappear when the game tree (and,
in particular, the information sets) are determined by the TM chosen, as
is the case in machine games.  Moreover, we show that,  
under natural assumptions regarding the complexity function, the two
notions coincide.

\fullv{The rest of this paper is organized as follows.  In
Section~\ref{sec:review}, we review the relevant definitions of Bayesian
machine game from \cite{HP08}.  In Section~\ref{sec:compex} we show
how we can view these Bayesian machine games as extensive-form games,
where the players moves involve computation.  In
Sections~\ref{sec:belief} we define beliefs in Bayesian machine games;
using this definition, we define interim and
ex ante sequential equilibrium in Section~\ref{sec:seqeqdef}, and
provide a natural condition under which they are equivalent.
In Section~\ref{sec:NEvsSE}, we relate Nash equilibrium
and sequential equilibrium.  Not surprisingly, every sequential
equilibrium in a Nash equilibrium; we provide a natural condition under
which a Nash equilibrium is an ex ante sequential equilibrium.   In
Section~\ref{sec:existence}, we consider when sequential equilibrium 
exists.  Since, as shown in \cite{HP08}, even Nash equilibrium may not
exist in Bayesian machine games, we clearly cannot expect a sequential
equilibrium to exist in general.  We show that if the set of TMs that
the agents can choose from is finite, then an ex ante sequential
equilibrium exists whenever a Nash equilibrium does; we also provide a
natural sufficient condition for an ex ante sequential
equilibrium to exist even if the set of TMs the agents can choose from
is infinite.  Up to this point in the paper, we have considered only
extensive-form games determined by Bayesian machine games, where players
make only one move in the underlying Bayesian game, and the remaining
moves correspond to computation steps.  In
Section~\ref{sec:extensiveform}, we extend our definitions to cases
where the underlying game is an extensive-form game.  We conclude in
Section~\ref{sec:discussion} with some discussion.}

\shortv{\vspace{-.1in}}
\section{Computational games: a review}\label{sec:review}
This review is largely taken from \cite{HP08}.   We model costly
computation using \emph{Bayesian machine games}.  Formally, a Bayesian
machine game is given by a tuple 
$([m], \M,T, \Pr,$ $\complexity_1, \ldots, \complexity_m,
u_1, \ldots, u_m)$, where
\beginsmall{itemize}
\item $[m] = \{1, \ldots, m\}$ is the set of players;
\item $\M$ is a set of TMs;
\item $\T \subseteq (\{0,1\}^{*})^m$ is the set of \emph{type 
profiles} ($m$-tuples consisting of one type for each of the $m$ players);%
\footnote{We have slightly simplified the definition in \cite{HP08}, 
by ignoring the type of nature, which gives
the formalism a little more power.  These changes are purely for ease of
exposition, to get across the main ideas.}
\item $\Pr$ is a distribution on $\T$;
\item $\complexity_i$ is a \emph{complexity function} (see below);  
\item $u_i: \T \times (\{0,1\}^*)^m \times \IN^m
\rightarrow \IR$ is player $i$'s utility function.   
Intuitively, $u_i(\vec{t},\vec{a},\vec{c})$ is the
utility of player $i$ if $\vec{t}$ is the type profile, $\vec{a}$ is the
action profile (where we identify $i$'s action with $M_i$'s output), and
$\vec{c}$ is the profile of machine complexities.  
\end{itemize}\vspace{-.1em}

If we ignore the complexity function and drop the requirement that an
agent's type is in $\{0,1\}^*$, then we have the standard definition of
a Bayesian game.  We assume that TMs take as input strings of 0s
and 1s and  output strings of 0s and 1s.  
Thus, we assume that both types and actions can be represented as
elements of $\{0,1\}^*$.  We allow machines to randomize, so given a
type as input, we actually get a distribution over strings.  
To capture this, we take the input to a TM to be not only a type,
but also a string chosen with uniform probability from 
$\{0,1\}^\infty$ (which we view as the outcome of an infinite
sequence of coin tosses).   
The TM's output is then a deterministic function of its type and
the infinite random string.
We use the convention that the output of a machine that does not
terminate is a fixed special symbol $\bott$.
We define a \emph{view} to be a pair $(t,r)$ 
of two bitstrings; we think of $t$ as that part of the
type that is read, and of $r$ as the string of random bits used. 
A complexity function $\complex: {\bf M}\times \bitset^*
; (\bitset^* \union \bitset^{\infty})
\rightarrow \IN$, where ${\bf M}$ denotes the set of Turing machines,
gives the complexity of a (TM,view) pair.

We can now define player $i$'s \emph{expected utility}
$U_i(\vec{M})$ if a profile $\vec{M}$
of TMs is played; we omit the standard details here.  
We then define a (computational) NE of a machine game
in the usual way:  
\BD Given a Bayesian machine game 
$G = ([m], \M,T, \Pr, \vec{\complexity},\vec{u})$,
a machine profile $\vec{M} \in \M^m$ is a \emph{(computational) Nash
equilibrium} if, for all players $i$, 
$U_i(\vec{M}) \ge U_i(M_i',\vec{M}_{-i})$ for all TMs $M_i' \in \M$.
\ED
Although a NE always exists in standard games, a
computational NE may not exist in 
machine games, as shown by the following example, taken from \cite{HP08}.
\begin{example} 
\label{roshambo}
{\rm Consider rock-paper-scissors.  As usual, rock beats scissors, scissors
beats paper, and paper beats rock. A player gets a payoff of 1 if he
wins, $-1$ if he loses, and 0 if it is a draw.  But now there is a
twist: 
since randomizing is cognitively difficult,
we charge players $\epsilon > 0$ for using a randomized strategy
(but do not charge for using a deterministic strategy).  Thus, a
player's payoff is $1-\epsilon$ if he beats the other player but uses a
randomized strategy.  
It is easy to see that
every strategy has a deterministic best response (namely, playing a best
response to whatever move of the other player has the highest probability); this is a
strict best response, since we now charge for randomizing.  It follows
that, in any equilibrium, both players must play deterministic
strategies (otherwise they would have a profitable deviation).  But
there is clearly no equilibrium where players use deterministic strategies.}
\bbox
\end{example}

Interestingly, it can also be shown that, in a precise sense, if there
is no cost for randomization, then a computational NE is
guaranteed to exist in a computable Bayesian machine game (i.e., one
where all the relevant probabilities are computable); see \cite{HP08}
for details.

\section{Computation as an extensive-form game}\label{sec:compex}
Recall that a deterministic TM $M=(\tau,\cQ,q_0,\cH)$ consists of a 
read-only input tape, a write-only output tape, a
read-write work tape, 3 machine heads (one reading each tape), a set
$\cQ$ of  machine states, a \emph{transition function} $\tau: \cQ \times 
\{0,1,b\}^2 \rightarrow \cQ \times \{0,1,b\}^2 \times \{L,R,S\}^3$, an
initial state $q_0 \in \cQ$, and a set $\cH\subseteq \cQ$ of ``halt'' states.
We assume that all the tapes are infinite, and that only 0s, 1s and blanks
(denoted $b$) are written on the tapes.  We think of the input to a
TM as a string in $\{0,1\}^*$ followed by blanks. 
Intuitively,
the transition function says what the TM will do if it 
is in a state $s$ and reads $i$ on the input tape and $j$ on the work
tape.  Specifically, $\tau$ describes what the new state is, what
symbol is written on the work tape and the output tape, and which way each of
the heads moves ($L$ for one step left, $R$ for one step right, or $S$
for staying in the same place).
The TM starts in state $q_0$, with the input written on the input tape,
the other tapes blank, the input head at the beginning of the input, and
the other heads at some canonical position on the tapes. 
The TM then continues computing according to $\tau$.
The computation ends if and when the machine reaches a 
halt state $q \in \cH$.
To simplify the presentation of our results, we restrict attention to
TMs that include only states $q \in \cQ$ that can be reached
from $q_0$ on some input. 
We also consider \emph{randomized} TMs, which are identical
to deterministic TMs except that the transition function $\tau$ now
maps $\cQ \times \{0,1,b\}^2$ to a probability distribution over $\cQ
\times \{0,1,b\}^2 
\times \{L,R,S\}^3$. 
As is standard in the computer science literature, we restrict attention
to probability distributions that
can be generated by tossing a fair coin; that is, the probability of all
outcomes has the form $c/2^k$ for $c, k \in \IN$.

In a standard extensive-form game, a strategy is a function from
information sets to actions.  Intuitively, in a state $s$ in an
information set $I$ for player $i$, the states in $I$ 
are the ones that $i$ considers possible, given his information at $s$;
moreover, at all the states in $I$, $i$ has the same information.  
In the ``runs-and-systems'' framework of
\fullv{Fagin et al.~\citeyear{FHMV},} 
\shortv{\cite{FHMV},}
each agent is in some \emph{local state} at 
each point in time.  
A \emph{protocol} for agent $i$ is a function from $i$'s local states to
actions.  We can associate with a local state $\ell$ for agent $i$ all the
histories of computation that end in that local state; this can be
thought of as the information set associated with $\ell$.
With this identification, a protocol can also be viewed as a
function from information sets to actions, just like a strategy.

In our setting, the
closest analogue to a local state in the runs-and-systems
framework is the state of the TM.  Intuitively, the TM's state describes
what the TM knows.%
\footnote{We could also take a local state to be the TM's state and the 
content of the tapes at the position of the heads.
However, taking the local state to be just the TM's state seems
conceptually simpler.  Taking the alternative definition of local state
would not affect our results.}
The transition function $\tau$ of the TM determines a protocol, that
is, a function from the TM's state to a ``generalized action'', 
consisting of reading the symbols on its read
and work tapes, then moving to a new state and writing some symbols on
the write and work tapes (perhaps depending on what was read).   We can
associate with a state $q$ of player $i$'s TM the information set $I_q$
consisting of all histories $h$ where player $i$'s is in state $q$ at the
end of the history.
(Here, a \emph{history} is just a sequence of
\emph{extended state profiles}, consisting of one
extended state for each player, and an \emph{extended state} for player $i$ 
consists of the TM that $i$ is using, the TM's state, and the content
and head position of each of $i$'s tapes.)
Thus, player $i$ implicitly \emph{chooses} his information sets (by
choosing the TM), rather than the information set being given exogenously. 
  
The extensive-form game defined by computation in a Bayesian
machine game is really just a collection of $m$ single-agent
decision problems.  
Of course, sequential equilibrium becomes even more interesting if we
consider computational extensive-form games, where there is computation
going on during the game, and we allow for interaction between the agents. 
\shortv{
We discuss sequential equilibrium in
(computational) extensive-form games in the full paper, which also
includes all proofs.%
\footnote{See http://www.cs.cornell.edu/home/halpern/papers/algseqeq.pdf.}
}
It turns out
that many of the issues of interest already arise in our setting.

\section{Beliefs}\label{sec:belief}
Using the view of machine games as extensive-form games,
we can define sequential equilibrium.  The first step to doing so
involves defining a player's 
beliefs at an information set.  But now we have to take into
account that the information set is determined by the TM chosen.
In the spirit of 
\fullv{Kreps and Wilson \citeyear{KW82},}  
\shortv{\cite{KW82},}  
define a \emph{belief system} $\mu$ 
for a game $G$ to be a function that associates with each player $i$, TM
$M_i$ for player $i$, and a state $q$ for $M_i$ a probability
on the histories in the information set $I_q$.
Following  \cite{HP09},  we interpret $\mu_{q,M_i}(x)$ as
the probability of 
going through
history $x$ conditional on 
reaching
the local state $q$.

We do not expect $\sum_{x \in I_q} \mu_{q, M_i} (x)$ to be $1$; in general, it
is greater than 1.  This point is perhaps best explained in the
context of games of imperfect recall.
Let the \emph{upper frontier} of an information set
$X$ in a game of imperfect recall, denoted $\hat{X}$, consist of all
those histories
$h \in X$ such that there is no history $h' \in X$ that is a prefix of
$h$.  In \cite{HP09}, we
consider belief systems that 
associate with each information set $X$ a 
probability $\mu_X$ on the histories in $X$.  Again,  we do not require
$\sum_{h \in X} \mu_X(h) = 1$.  For example, if all the histories in $X$
are prefixes of same complete history $h^*$, then 
we might have $\mu_X(h) = 1$ for all
histories $h \in X$.  However, we do require that $\Sigma_{h \in \hat{X}}
\mu_X(h) = 1$.  We make the analogous requirement here. Let
$\hat{I}_q$ denote the upper frontier of $I_q$.
The following lemma, essentially already proved in \cite{HP09},
justifies the requirement.
\begin{lemma}
If 
$q$ is a local state for player $i$ that is reached by $\vec{M}$ with
positive probability, and $\mu_q'(h)$ is the probability of  going
through history
$h$ when running $\vec{M}$ conditional on reaching $q$, then 
$\sum_{h \in \hat{I}_q} \mu_q'(h) = 1$.
\end{lemma}

Given a belief system $\mu$ and a machine profile $\vec{M}$, define a
probability distribution $\mu^{\vec{M}}_q$ over terminal 
histories in the obvious way: for each terminal history $z$, 
let $h_z$ be the history in $\hat{I}_q$ generated by
$\vec{M}$ that is a prefix of $z$ if there is one (there is clearly at
most one), 
and define  $\mu^{\vec{M}}_q(z)$ as the product of $\mu_{q,M_i}(h_z)$
and the probability that $\vec{M}$ leads to the terminal history $z$
when started in 
$h_z$; if there is no prefix of $z$ in $I_q$, then $\mu^{\vec{M}}_q(z) =
0$. 
\fullv{Following Kreps and Wilson \citeyear{KW82},} 
\shortv{Following \cite{KW82},} 
let $U_i(\vec{M} \mid q,\mu)$
denote the expected utility for player $i$, where the expectation is taken
with respect to $\mu^{\vec{M}}_q$. 
Note that utility is well-defined, since  a terminal history
determines both the input and the random-coin flips of each player's
TM, and thus determines both its output and complexity.

\section{Defining sequential equilibrium}\label{sec:seqeqdef}
If $\Qchange$ is a state of TM $M_i$,
we want to capture the intuition that a TM $M_i$ 
for player $i$ is a best response
to a machine profile $\vec{M}_{-i}$ for the remaining players
at $q$ (given beliefs $\mu$).
Roughly speaking, we capture this by requiring that the expected utility
of using another TM $M'_i$ starting at a node where the TM's state is
$\Qchange$ to be no 
greater than  that of using $M_i$.   
``Using a TM $M'_i$ starting from $\Qchange$'' means
using the TM $(M_i,\Qchange,M_i')$, which, roughly speaking, is the TM
that runs like $M_i$ up to $\Qchange$, and then runs like $M'_i$.%
\footnote{We remark that, in the definition of sequential equilibrium in
\cite{HP09}, the agent was allowed to change strategy not just at a
single information set, but at a collection of information sets.  
Allowing changes at a set of information sets seems reasonable for an ex
ante notion of sequential equilibrium, but not for an interim notion;
thus, we consider changes only at a single state here.
\fullv{
Allowing changes at a set of states here, the analogue of what was done
in \cite{HP09} would give a refinement of our
definition (i.e., we would have fewer sequential equilibria), but all
our basic results would hold with the proofs essentially unchanged.
One reason for considering a set of information sets in \cite{HP09} was
to ensure that every sequential equilibrium is a NE.  As
we shall see, we already have that property in the computational
setting.}}
\commentout{Here's the example, which we may want to include in the full
paper: Suppose that there are two types: 0 has probability 1, and 1
has probability 0.  The payoffs if the input is t1 are 
10 if the output is 0, 1 if the output 11, 5 if the output is 00, and 0
if the output is 10 or 01.  On input 0, TM M writes 0, moves to state
q1 and halts.  On input 1, M randomizes, moving to each of q2 and q3
with probability 1/2.  In q2, if the work tape is blank, it writes 0 on
the output tape, 1 on the worktape, and transitions to q3; in q3, if the
work tape is blank, it writes 0 on the output tape, 1 on the worktape,
and transitions to q2.   In q2 or q3, if there is a 1 on the work tape, it
write 0 on the output tape, transitions to q1, and halts.  Since q3 is
not below q2, nor is q2 below q3, it is easy to see that this is a
sequential equilibrium in the weak sense.  The player does not want to
switch at either q2 or q3.  But the player would want to change
conditional reaching {q2,q3}.   
}

In the standard 
setting, all the subtleties in the definition of sequential equilibrium 
involve dealing 
with what happens at information sets that are reached with probability 0.
When we consider machine games, as we shall see, under some reasonable
assumptions, all states are reached with positive probability in
equilibrium, so dealing with probability 0 is not a major concern (and,
in any case, the same techniques that are used in the standard setting
can be applied). But there are several new issues that must be addressed
in making precise what it means to ``switch'' from $M_i$ to $M_i'$ at
$\Qchange$.

Given a TM $M_i = 
(\tau,\cQ,q_0,\cH)$ and $\Qchange \in \cQ$, let 
$\cQ_{\Qchange,M_i}$ consist of all states $q' \in \cQ$ such 
that, for all 
views $v$,
if the computation of $M_i$ given
view $v$
reaches $q'$, then it also reaches $\Qchange$. 
We can think of $\cQ_{\Qchange,M_i}$ as consisting
of the states $q'$ that are ``necessarily'' below $\Qchange$, given that 
$M_i$ is used. Note that $\Qchange \in \cQ_{\Qchange,{M_i}}$.%
\footnote{Note that in our definition of $\cQ_{\Qchange,M_i}$ (which defines what
``necessarily'' below $\Qchange$ means) we consider all 
possible computation
paths, and in particular also paths that are not reached in
equilibrium. An alternative definition would consider only paths of
$M_i$ that are reached 
given the particular type distribution.
We make the former choice since
it is more consistent with the traditional presentation of sequential
equilibrium (where histories off the equilibrium path play an important
role).}
Say that $M_i' = (\tau',\cQ',q',\cH')$ is \emph{compatible with $M_i$ given
$\Qchange$} if 
$q' = q$ (so that $q$ is the start state of $M_i'$) and
$\tau$ and $\tau'$ agree on all states in $\cQ-
\cQ_{\Qchange,M_i}$. 
If $M_i'$ is not compatible with $M_i$ given $\Qchange$, then
$(M_i,\Qchange,M_i')$ is not well defined.  If $M_i'$ is
compatible with $M_i$ given $\Qchange$, then $(M_i,\Qchange,M_i')$ is
the TM 
$(\cQ'', [\tau,\Qchange,\tau'], q_0,\cH'')$, where 
$\cQ'' = (\cQ - \cQ_{\Qchange,M_i}) \cup \cQ'$;
$[\tau ,\Qchange, \tau']$ is the  
transition function that agrees with $\tau$ on states in $\cQ -
\cQ_{\Qchange,M_i}$, and agrees with $\tau'$ on the remaining states; and
$\cH'' = (\cH \cap (\cQ - \cQ_{\Qchange,M_i})) \cup \cH'$.

Since $(M_i,\Qchange,M_i')$ is a TM, its complexity given a view is well
defined. 
In general, the complexity of $(M_i,\Qchange,M_i')$
may be different from that of $M_i$ even on histories that do not go
through $q$.  For example, consider a one-person decision problem where
the agent has an input (i.e., type) of either 0 or 1.  Consider four
TMs: $M$, $M_0$, $M_1,$ and $M^*$.  Suppose that $M$ embodies a simple
heuristic that works well on both inputs, $M_0$ and $M_1$ give
better results than $M$ if the inputs are 0 and 1,  
respectively, and $M^*$ acts like $M_0$ if the input is 0
and like $M_1$ if the input is 1.  Clearly, if we do not take
computational costs into account, $M^*$ is a better choice than $M$;
however, suppose that with computational costs considered, $M$ is better
than $M_0$, $M_1$, and $M^*$.  Specifically, suppose that $M_0$, $M_1$,
and $M^*$ all use more states than $M$, and the complexity function
charges for the number of states used.  
Now suppose that the agent moves to state
$q$ if he gets an input of 0.  In state $q$, using $M_0$ is
better than continuing with $M$: the extra charge for complexity is
outweighed by the improvement in performance.  Should we say that using
$M$ is then not a 
sequential equilibrium?  The TM $(M,\{q\},M_0)$ acts just like $M_0$ if the
input is 0.  
From the ex ante point of view, $M$ is a better
choice
than $M_0$.  However,  having reached $q$, the 
agent arguably does not care about the complexity of $M_0$ on
input 1.  Our definition of ex ante sequential equilibrium
restricts the agent to making changes that leave unchanged the
complexity of paths  that do not go through $q$ (and thus would not
allow a change to $M_0$ at $q$). Our definition of \emph{interim}
sequential equilibrium does not make this restriction; this is the only
way that the two definitions differ.%
Note that this makes it easier for a strategy to be an ex ante sequential
equilibrium (since fewer deviations are  considered).  

$(M_i,\Qchange,M_i')$ is a \emph{local variant of $M_i$} if the complexity of
$(M_i,\Qchange,M_i')$ is the same as that of $M_i$ on views that do not go
through $\Qchange$; that is, if for every view $v$
such that the computation of $M_i(v)$ does not reach $\Qchange$,
$\complex(M_i,v) = \complex((M_i,\Qchange,M_i'),v)$.
A complexity function $\complex$ is \emph{local} if 
$\complex(M_i,v) = \complex((M_i,\Qchange,M_i'),v)$ for all 
TMs $M_i$ and $M_i'$, states $\Qchange$, and views $v$ that do not reach
$\Qchange$. 
Clearly a complexity function that considers only running time, space used, and
the transitions undertaken on a path is local.
If $\complex$ also takes the number of states into account, then
it is local as long as $M_i$ and 
$(M_i,\Qchange,M_i')$ have the same number of states.
Indeed, if we think of the state space as ``hardware'' and the
transition function as the ``software'' of a TM, then restricting to
changes $M_i'$ that have the same  
state space as $M_i$ seems reasonable: when the agent contemplates
making a change at a non-initial state, he cannot 
acquire new hardware, so he must work with his current hardware.

A TM $M_i = (\tau,\cQ,q_0,\cH)$ for player $i$ is \emph{completely
mixed} 
if, for all states $q \in \cQ - \cH$, $q' \in \cQ$, and bits $k, k' \in
\{0,1\}$, 
$\tau(q,k,k')$ assigns positive 
probability to making a transition to $q'$.
A machine profile $\vec{M}$ is completely
mixed if for each player $i$, $M_i$ is completely mixed. 
\fullv{Following Kreps and Wilson \citeyear{KW82},} 
\shortv{Following \cite{KW82},} 
we would like to say that a
belief  
system $\mu$ is \emph{compatible} with a machine profile $\vec{M}$ 
if there exists a sequence of completely-mixed machine profiles $\vec{M^1},
\vec{M^2}, \ldots$ converging to $\vec{M}$ such that 
if $q$ is a local state for player $i$ that is reached with positive
probability by $\vec{M}$ (that is, there is a type profile $\vec{t}$
that has positive probability according to the type distribution in $G$
and a profile $\vec{r}$ of random strings such that
$\vec{M}(\vec{t},\vec{r})$ reaches $q$), then $\mu_{q,M_i}(h)$ is just
the probability of $\vec{M}$ going through $h$ conditional on $\vec{M}$
reaching $q$ (denoted 
$\pi_{\vec{M}}(h \mid q)$); and if $q$ is a 
local state that is reached with probability 0 by $\vec{M}$, then
$\mu_{q,M_i}(h)$ is $\lim_{n \rightarrow \infty} \pi_{\vec{M}^n}(h \mid q)$.
To make this precise, we have to define convergence.
We say that 
\emph{$\vec{M^1}, \vec{M^2}, \ldots$ converges to $\vec{M}$} if, for
each player $i$, 
all the TMs $M^1_i,{M^2_i}, \ldots, {M_i}$
have the same state space, and 
the transition function of each TM in the
sequence to converge to that of $M_i$.  
Note that we place no requirement on the complexity functions.  We could
require the complexity function of $M_i^k$ to converge to that of $M_i$
in some reasonable sense.  However, this seems to us unreasonable.  
\fullv{If we assume that randomization is free (in the sense 
hinted at after Example~\ref{roshambo}), then the convergence of the
complexity 
functions follows from the convergence of the transition functions.
On the other hand,}
\shortv{For example,} 
if we have a complexity function that charges for
randomization, as in Example~\ref{roshambo}, then the complexity
functions of $M_i^n$ may not converge to the complexity function of $M_i$.
Thus, if we require the complexity
functions to converge, there will not be a sequence of completely mixed
strategy profiles converging to a deterministic strategy profile
$\vec{M}$.  
If we think of the sequence of TMs as arising from 
``trembles'' in the operation of some fixed TM (e.g., due to
machine failure), then requiring that the complexity functions converge
seems unreasonable.  

\BD 
\label{seqeq.def}
A pair $(\vec{M},\mu)$ consisting of a machine profile $\vec{M}$ and
a belief system 
$\mu$ is called a \emph{belief assessment}.  A belief assessment
$(\vec{M},\mu)$ is an \emph{interim sequential equilibrium} (resp.,
\emph{ex ante sequential equilibrium}) in 
a machine game
$G = ([m],\M,\ldots)$ if $\mu$ is compatible with $\vec{M}$ and
for all players $i$, states $\Qchange$  of $M_i$,
and TMs $M'_i$ compatible with $M_i$ and $\Qchange$
such that 
$(M_i,\Qchange,M'_i) \in \M$ (resp., and 
$(M_i,\Qchange,M'_i)$ is a local variant of $M_i$),
we have $$U_i(\vec{M} \mid q,\mu) \ge
U_i(((M_i,\Qchange,M'_i),\vec{M}_{-i}) \mid 
q, \mu)).$$  
\ED
\fullv{Note that in}
\shortv{In} 
Definition \ref{seqeq.def} we consider only switches
$(M_i,\Qchange,M'_i)$ that result in a TM  that is in the set
$\M$ of possible TMs in the game.  
That is, we require that the TM we switch to is ``legal'', and has a
well-defined complexity.

As we said, upon reaching a state $q$, an agent may well want to switch
to a TM $(M_i,\Qchange,M'_i)$ that is not a local variant of $M_i$.
This is why we drop this requirement in the definition of interim
sequential equilibrium. But it is a reasonable requirement ex ante.  It 
means that, at the \emph{planning stage of the game}, there is no
TM $M_i'$ and state $q$ such that agent $i$ prefers to
use $M'_i$ in the event that $\Qchange$ is reached.   
That is, $M_i$ is ``optimal'' at the planning stage,
even if the agent considers the possibility of reaching states 
that are off the equilibrium path. 

The following result is immediate from the definitions, and shows that,
in many cases of interest, the two notions of sequential equilibrium coincide.
\begin{proposition}\label{pro:obvious} Every interim sequential
equilibrium is an ex ante sequential equilibrium.  In a machine game
with a local complexity function, the interim and ex ante sequential
equilibria coincide.
\end{proposition}

\fullv{As the discussion above emphasizes,
a host of new issues arise when defining
sequential equilibrium in the context of machine games.  While we
believe we have made reasonable choices, variants of our definitions are
also worth considering.}
For instance, our way of defining beliefs in the definition of interim
  sequential equilibrium arguably still has an ex ante
  flavor (recall that the probability assigned to a node in an
  information set is the
  probability of \emph{reaching} the node conditioned on \emph{reaching} the
  information set). If we want these beliefs to instead
  correspond to the probabilities the agent assigns to \emph{being
    at} the node, conditioned on \emph{being at} the information set,
we need to consider more carefully when the agent 
reconsiders his strategy.  If the decision to reconsider depends only on
the information set, then the change in strategy must happen at the
upper frontier of the information set, that is, when the information set
is first reached, and our current analysis of interim sequential seems
reasonable.  If the change does not necessarily happen at the upper
frontier, then we need to model under what circumstances reconsideration
occurs.  (This point is also made in \cite{GroveHalpern95,Hal15}.)
For concreteness, in our context, 
assume that at each step of the computation, with some small
  probability $\epsilon$, a human (the agent) comes in, observes
  the state of the TM, and decides whether it wishes to switch
  machines\footnote{Even fully specifying such a model requires some
    care. For instance, can the agent come in twice? 
In our discussion, for definiteness, we assume that 
    the agent comes in only once.}

Such a model would result in a different way of ascribing
  beliefs. Our results no longer apply if we use this alternative way
  of ascribing beliefs: it is not hard to come up with a machine game
that corresponds to the ``absentminded-driver'' game of \cite{PR97}
where a Nash equilibrium exists, the complexity function is local, but
no interim sequential equilibrium exists using this method to ascribe beliefs.

\section{Relating Nash equilibrium and sequential
equilibrium}\label{sec:NEvsSE} 
In this section, we relate NE and sequential equilibrium.

First note that is easy to see that every (computational) 
sequential equilibrium is a NE,
since if $\Qchange = q_0$, the
start state, $(M_i,q_0,M_i') = M_i'$.  That is, by taking $\Qchange =
q_0$, we can consider arbitrary modifications of the TM $M_i$.%
\footnote{As we mentioned earlier, since a game tree was not assumed to
have a unique initial node 
in \cite{HP09}, it was necessary to allow changes at sets of information
sets to ensure that every sequential equilibrium was a NE.}

\begin{proposition}\label{seqeqisNash} Every ex ante sequential
equilibrium is a NE. 
\end{proposition}
\fullv{Of course, since} \shortv{Since} every interim sequential
equilibrium is an ex ante 
sequential equilibrium, it follows that every interim sequential
equilibrium is a NE as well.

In general, not every NE is a sequential equilibrium.
However, under some natural assumptions on the complexity function, the
statement holds.
A strategy profile $\vec{M}$ in a machine game $G$ is
\emph{lean} if, for all 
players $i$ and 
local states $q$ of $M_i$, $q$ is reached with positive probability when
playing $\vec{M}$.  
\fullv{The following proposition is the analogue of the well-known
result that, with the traditional definition of sequential equilibrium, every
completely mixed NE is also a sequential equilibrium. }

\begin{proposition}\label{lem:nashmixed0}
If $\vec{M}$ is a lean NE for 
machine
game $G$  
and $\mu$ is a belief system compatible with $\vec{M}$, then $(\vec{M},
\mu)$ is an ex ante sequential equilibrium.%
\end{proposition}

\fullv{
\BPRF
We need to show only that for each player $i$ and 
local state $q$
of $M_i$, there does not exist a TM $M'_i$ compatible with $(M_i,q)$
such that $U_i(\vec{M} \mid q,\mu) < U_i(((M_i,q,M'_i),\vec{M}_{-i})
\mid q, \mu).$ 
Suppose by way of contradiction that there exist such a TM $M'_i$
and local state $q$. 
Since $\mu$ is compatible with $\vec{M}$, it follows that  
$U_i(\vec{M} \mid q) < U_i(((M_i,q,M'_i),\vec{M}_{-i}) \mid q).$
Since $(M_i,q,M_i')$ is local variant of $M_i$,
$U_i(\vec{M} \mid  \text{not reaching } q) =
U_i(((M_i,q,M'_i),\vec{M}_{-i}) \mid \text{not reaching } q)).$ 
By the definition of $(M_i,q,M'_i)$, the probability that
$M_i$ and $(M_i,q,M'_i)$ reach $q$ is identical; it follows that
$U_i(\vec{M}) < U_i((M_i,q,M'_i),\vec{M}_{-i})$, which contradicts the
assumption that $\vec{M}$ is a NE. 
\EPRF
\bigskip
}

The restriction to local variants $(M_i,\Qchange,M_i')$ of $M_i$ in the
definition of ex ante sequential equilibrium is critical here.
Proposition~\ref{lem:nashmixed0} does not hold for interim sequential
\shortv{equilibrium, as the following example, which illustrates the role of
locality, shows.}
\fullv{equilibrium.  Suppose, for example, that 
if $i$ is willing to put in more computation at $q$, then he gets a
better result.  Looked at from the beginning of the game, it is not
worth putting in the extra computation, since it involves using extra
states, and this charge is global (that is, it affects the complexity of
histories that do not reach $q$).  But once $q$ is reached, it is
certainly worth putting in the extra computation.  If we
assume locality, then the extra computational effort at $q$ does not
affect the costs for histories that do not go through $q$.  Thus, if it
is worth putting in the effort, it will be judged worthwhile ex ante.
The following two examples illustrate the role of locality.}

\begin{example}\label{xam:incompressible}
{\rm Let $x$ be an $n$-bit string whose Kolmogorov complexity is
$n$ (i.e., $x$ is incompressible---there is no shorter description of
$x$). Consider a single-agent game
$G_x$ (so the $x$ is built into the game; it is not part of the input)
where the agent's type is a string of length $\log n$, chosen uniformly at
random, and the utility function is defined as follows, for an agent
with type $t$:
\begin{itemize}
\item The agent ``wins'' if it outputs $(1,y)$, where $y = x_t$ (i.e.,
it manages to guess the $t$th bit of $x$, where $t$ is its type). In
this case, it receives a 
utility of $10$, as long as its complexity is at most 2. 
\item The agent can also ``give up'': if it outputs $t_0$ (i.e., the
first bit of its type) and its complexity is 1, then it receives a
utility of 0. 
\item Otherwise, its utility is $-\infty$.%
\footnote{We can replace
$-\infty$ here by any sufficiently small integer;  $-20\log n$ will do.}
\end{itemize}
Consider the 4-state TM $M$ that just ``gives up''.  Formally, 
$M = (\tau,\{q_0,b_0,b_1,H\},q_0,\{H\})$, where $\tau$ is such that, 
in $q_0$, $M$ reads the
first bit $t_0$ of the type, and transitions to $b_i$ if it is $i$; 
and in state $b_i$, it outputs $i$ and transitions to $H$, the halt state.
\fullv{
Now define the complexity function as follows:
\begin{itemize}
\item the complexity of $M$ is $1$ (on all inputs);
\item the complexity of any TM $M' \neq M$  that has at most $0.9n$
states is $2$; 
\item all other TMs have complexity 3.
\end{itemize}
}}

\shortv{Take the complexity of $M$ to be $1$ (on all inputs);
the complexity of any TM $M' \neq M$  that has at most $0.9n$
states is $2$; and all other TMs have complexity 3.}

{\rm Note that $M$ is the unique NE in $G_x$.
Since $x$ is incompressible, no TM $M^*$ with fewer than $0.9n$
states can correctly guess $x_t$ for all $t$ (for otherwise $M^*$ would
provide a description of $x$ shorter than $|x|$).   It follows that no TM
with complexity greater than 1 does better than $M$. Thus, $M$ is the
unique NE.
It is also a lean NE, and thus, by Proposition
\ref{lem:nashmixed0}, an ex ante sequential equilibrium. 
However, there exists a non-local variant of $M$ at $b_0$ that gives higher
utility than
$M$.  Notice that if the first bit is 0 (i.e., if the TM is in
state $b_0$), then $x_t$ is one of the first $n/2$ bits of $x$. 
Thus, at $b_0$, we can switch to the TM $M'$ that reads the whole
type $t$ 
and the first $n/2$ bits of $x$, outputs $(1,x_t)$. It is easy to see
that $M'$ can be constructed using $0.5n+O(1)$ states.  Thus, $M$ is not
an interim sequential equilibrium (in fact, none exists in $G_x$).
$(M,b_0,M')$ is not a local variant of $M$, since $M'$ has higher
complexity than $M$ at $q_0$.  }
\qed
\end{example}

\begin{example} 
{\rm Consider the game in Figure~\ref{fig2} again.  Recall that 
the strategy that maximizes expected utility is the strategy $f$ that
chooses action $S$ at node $x_1$, action $B$ at node
$x_2$, and action $R$ at the information set $X$ consisting of $x_3$ and
$x_4$.  Let $f'$ be the strategy of choosing
action $B$ at $x_1$, action $S$ at $x_2$, and $L$ at $X$.
As Piccione and Rubinstein point out, if node $x_1$ is reached and the
agent is using $f$, then he will not want to continue using $f$; he
would prefer to switch to $f'$ instead.  In the
language of this paper, $f$ is not an interim sequential equilibrium,
although it is a NE of the one-player game.  Note that
$f'$ is neither a NE nor an interim sequential equilibrium
(since if the player is using $f'$ at $x_2$, he will want to switch to
$f$).   According to the definition in \cite{HP09}, $f$ is an ex
ante sequential equilibrium.   The reason is that switching from $f$ to
$f'$ is not allowed at $x_1$, because $f'$ does something different from
$f$ at a node that is not below $x_1$, namely $x_4$.  As a consequence,
$(f,x,f')$ is not a strategy in the game, since it does different things
at $x_3$ and $x_4$, although they are in the same information set.  The
requirement made in \cite{HP09} that, when considering switching from a
strategy $f$ to $f'$ at an 
information set $X^*$, $f'$ has to agree with $f$ at all nodes not below
$X^*$ is somewhat analogous to the 
local-variant requirement that we make here in the
definition of ex ante sequential equilibrium.}

{\rm We now consider a machine game that captures some of the essential
features of the game in Figure~\ref{fig2}.  Suppose that there are two
types, 0 and 1, which each occurs with probability $1/2$.  The agent must
choose between two TMs, $M_0$ and $M_1$, which can be viewed as
corresponding to $f$ and $(f,x,f')$.   $M_0$ reads the input in state
$q_0$, and moves to either state $q_1$ or $q_2$, depending on whether it
reads 0 or 1.  In $q_1$, $M_0$ moves to state $H$, the halt state,
outputting nothing (i.e., the output is $b$, the blank symbol).  In
state $q_2$, $M_0$ writes 1 on its output tape and moves to $q_4$; 
in $q_4$, it writes 1 again, and moves to $H$.
$M_1$ again moves to $q_1$ or $q_2$ depending on its input, and from
$q_2$ moves to $q_4$ and $H$, writing $11$, just like $M_0$.  But from
$q_2$, it moves to $q_3$ and $H$, writing $00$.  }

{\rm The payoffs are as follows: if the output is $b$, the payoff is $2-c$
for both inputs, where $c$ is the complexity of the machine chosen.  If
the output is $00$ and the input is $0$, the payoff is $3-c$; if the 
output is $11$ and the input is $1$, the payoff is $4-c$.  If 
the output is something other than $b$ or $00$ and the input is 0, then the
payoff is $-6$; if the output is something other than $b$ or
$11$ and the input is 1, then the payoff of $-2$.  Finally, $c = 0$ if
$M_0$ is chosen, and $c=.75$ if $M_1$ is chosen.}

{\rm It is easy to see that  $M_0$ is a NE; its expected
payoff is 3, while that of $M_1$ is $2.75$. However, $M_0$ is not an
interim sequential equilibrium, because at $q_1$, the agent prefers
switching 
to $(M_0,q_1,M_1)$, which is equivalent to $M_1$, since,
conditional on reaching $q_1$, the expected payoff of $M_0$ is $2$,
while that of $M_1$ is $2.25$.   Note that although $(f,x_1,f')$ is not
a strategy, $(M_0,q_1,M_1)$ is a TM (and is equivalent to $M_1$).
Switching from $M_0$ to $(M_0,q_1,M_1)$ results in changing the
information strucure.  Such a change is not possible in standard games
of imperfect recall.  Finally, note that $M_0$ is an ex ante sequential
equilibrium; the switch from $M_0$ to $(M_0,q_1,M_1)$ is disallowed at
$q_1$ because $(M_0,q_1,M_1)$ is not a local variant of $M_0$.}
\qed
\end{example}

We now show that for a natural class of games, every NE is
lean, and thus also an ex ante sequential equilibrium. 
Intuitively, we
consider games where there is a strictly positive cost for having more
states. 
Our argument is similar in spirit to that of 
\fullv{Rubinstein \citeyear{Rub85}, who showed that in his games with
automata,} 
\shortv{\cite{Rub85}, where it is shown that in games with automata,} 
there is
always a NE with no ``wasted'' states; all states are
reached in equilibrium.
Roughly speaking, a machine game $G$ has \emph{positive state
cost} if (a) a state $q$ is not reached in TM $M$, and $M^{-q}$ is the
TM that results from removing $q$ from $M$, then $\complex_i(M^{-q},v) <
\complex_i(M,v)$; and (b)  utilities are monotone decreasing in
complexity; that 
is, $u_i(\vec{t}, \vec{a}, (c'_{i},\vec{c}_{-i})) < u_i(\vec{t}, \vec{a},
(c_{i},\vec{c}_{i}))$ if $c'_i > c_i$.
\shortv{(See the full paper for the precise definition.)}
\fullv{More precisely, we have the following definition.}

\fullv{
\BD \label{monotone.def}
A machine
game $G=([m], \M,\Pr,\vec{\complex},\vec{u})$ has \emph{positive
state cost} if the following two conditions hold: 
\begin{itemize}
\item For all players $i$, TMs $M_i = (\tau,\cQ,q_0,\cH)$, views $v$ for player
$i$, and  
local states $q \ne q_0$ in $\cQ$
such that $q$ is not reached  in view $v$ when running $M_i$ (note
that because the view gives the complete history of messages received and
read, we can compute the sequence of states that player $i$ goes through
when using $M_i$ if his view is $v$),
$\complex_i(M^{-q},v) < \complex_i(M,v)$, where $M^{-q} = (\cQ - \{q\},
\tau^q, q_0)$, and $\tau^q$ is identical to $\tau$ except that all
transition to $q$ are replaced by transitions to $q_0$.
\item 
Utilities are \emph{monotone decreasing in complexity}; that is, 
for all players $i$, type profiles $\vec{t}$, action profiles
$\vec{a}$, and complexity profiles $(c_{i},\vec{c}_{-i})$,
views $v$,
$(c'_{i},\vec{c}_{-i})$, if $c'_{i} > c_{i}$, then $\complex_i(M^{-q},v)
< \complex_i(M,v)$.
\end{itemize}
\end{dfn}
}

\BL\label{lem:nashmixed} Every NE $\vec{M}$ for 
machine game $G$ with positive 
state cost is lean.
\EL
\fullv{
\BPRF 
Suppose, by way of contradiction, that there exists a NE
$\vec{M}$ for a game $G$ with positive state cost, a player $i$,
and a local state $q$ of $M_i$ that is reached with probability $0$.
First, note that $q$ cannot be the initial state of $M_i$, 
since, by definition, the initial state of every TM is reached with
probability 1.
Since $g$ has positive state cost, for
every view $v$ that is assigned positive probability (according to the type
distribution of $G$), $M_i^{-q}(v)$ has the same output as $M_i(v)$ and
$\complex_i(M_i^{-q},v)< \complex(M_i^{-q},v)$. 
Since the utility is monotonic in complexity, it follows that
$U_i(M^{-q}_i,\vec{M}_{-i}) < U_i(\vec{M})$, which contradicts 
the assumption
that $\vec{M}$ is a NE.
\EPRF
}

Combining Proposition~\ref{lem:nashmixed0} and Lemma~\ref{lem:nashmixed}, we
immediately get the following result.
\BT 
\label{nashimpliesseq.thm}
If $\vec{M}$ is a NE for a
\mbox{machine game $G$} with
positive state cost, and $\mu$ is a belief system compatible \mbox{with
$\vec{M}$, then $(\vec{M}, \mu)$ is an ex ante sequential equilibrium.}
\ET

One might be tempted to conclude from Theorem \ref{nashimpliesseq.thm}
that sequential equilibria are not interesting, since every NE
is a sequential equilibrium. 
But this result depends on the assumption of positive state cost in a
critical way, as
the following 
simple 
example shows.

\begin{example}\label{poscost.ex}
{\em
Consider a single-agent game where the type space is $\{0,1\}$,
and the agent gets payoff 1 if he outputs his type, and otherwise gets 0.
Suppose that all TMs have complexity 0 on all inputs (so that the game
does not have positive state cost),
and that the type distribution assigns probability 1 to the type being 0.
Let $M$ be the 4-state TM that reads the input and then outputs 0.
Formally, $M = (\tau,\{q_0,b^0,b^1,H\},q_0,\{H\})$, where $\tau$ is such that 
in $q_0$, $M$ reads the type $t$ and transitions to $b_i$ if the type is $i$;
and in state $b_i$, it outputs $0$ and transitions to $H$, the halt state.
$M$ is clearly a NE, since $b=0$ with probability
1. However, $M$ is not an ex ante sequential equilibrium, since conditioned on
reaching $b_1$, outputting 1 and transitioning to $H$ yields higher
utility; furthermore, note that this change is a local variant of $M$ since
all TMs have complexity 0.
} 
\qed
\end{example}

\commentout{
In standard games, every sequential equilibrium is a Nash equilibrium.
We now shown that, under a minimal assumption on the complexity
function, the same is true in machine games.
A complexity function $\complex$ is \emph{invariant under
isomorphism} if, for all TMs $M$ and $M'$ that are identical up
to a renaming of the state space and every view $v$,
$\complex(M,v)=\complex(M',v)$.%
\footnote{Arguably, invariance under isomorphism is a property we should
have required of complexity functions all along.}
\BT \label{seqimpliesnash.thm} If  $(\vec{M}, \mu)$ is a sequential
equilibrium for a machine game $G$ whose complexity functions are
invariant under isomorphism, then $\vec{M}$ is a NE for
$G$. 
\ET
\BPRF
Suppose that  $(\vec{M}, \mu)$ is a sequential equilibrium of $G$, but
$\vec{M}$ is not a NE.  Then there exists some player $i$
and a TM $M'_i$ such that $U_i(M_i',\vec{M}_{-i}) > U_i(\vec{M})$.
By invariance under isomorphism, we can assume without
loss of generality that $M$ and $M'$ have the same start state $q_0$. 
Since, by assumption, we consider only TMs where all states can be
reached from the start state, it follows that
$(M_i,q_0,M'_i)=M'_i$, and that 
$$U_i(((M_i,q_0,M'_i),\vec{M}_{-i}) \mid 
q_0, \mu)) > U_i(\vec{M} \mid q_0,\mu).$$
$(M_i,q_0,M'_i)$ is trivially a local variant of $M_i$ (since there is
no view $v$ such that the computation of $M_i(v)$ does not reach
$q_0$).  This gives us a contradiction to the assumption that
$(\vec{M},\mu)$ is a sequential equilibrium.
\EPRF
}

\fullv{
In the next example, we speculate on how Theorem
\ref{nashimpliesseq.thm} can be applied to reconcile causal determinism
and free will. 
\begin{example} [Reconciling determinism and free will]
{\em
\emph{Bio-environmental determinism} is the idea that all our behavior is
determined by our genetic endowment and the stimulus we receive. 
In other words, our DNA can be viewed as a program (i.e., a TM) 
which acts on the input signals that we receive.
We wish to reconcile this view with the idea that people have a feeling of free will, and more precisely, that people have a feeling of actively making (optimal) decisions.}

{\em Assume that our DNA sequences encode a TM such that the profile of
TMs is a NE of a Bayesian
machine game $G$ (intuitively, the ``game of life''). Furthermore, assume
that the states of that TM
correspond to the ``conscious'' states of computation. That is,
the states of the TM consists only of states that intuitively correspond
to conscious moments of decision; all subconscious computational
steps are bundled together into the transition function. 
If the game $G$ has positive state cost then, by Theorem
\ref{nashimpliesseq.thm}, we have a sequential equilibrium, so at each 
``conscious state'', an 
agent does not want to change its program. In other words, the agent
``feels'' that its action is optimal.} 

{\em An ``energy argument'' could justify the assumption that $G$ has
positive state cost: if two DNA sequences encode exactly the same
function, but the program describing one of the sequences has less
states than the other then, intuitively, only the DNA sequence encoding
the smaller program ought to survive evolution---the larger program
requires more ``energy'' and is thus more costly for the organism. In
other words, states are costly in $G$. As a result, we have that agents
act optimally at each conscious decision point.
Thus, although agents feel that they have the option of changing their
decisions at the conscious states, they choose not to.
} 
\qed
\end{example}
}

\section{Existence}\label{sec:existence}
We cannot hope to prove a general existence theorem for sequential
equilibrium, since not every game has even a NE, and 
by Proposition \ref{seqeqisNash},
every ex ante sequential equilibrium is a NE. 
Nonetheless, we show that 
for any Bayesian machine game $G$ where the set $\M$ of possible TMs
that can be chosen is finite, if $G$ has a NE, 
then it has a sequential equilibrium. 
More precisely, we show that 
in every game where the set $\M$ of possible TMs that can be chosen
is finite, 
every NE can be converted to
an ex ante sequential equilibrium with the same distribution over outcomes. 
As illustrated in Example \ref{poscost.ex}, not every NE is 
an ex ante sequential equilibrium;
thus, in general, we must modify the original
equilibrium. 

\begin{thm}
Let $G$ be a machine game where the set $\M$ of TMs is finite.
If 
$G$ has a NE, then it has an ex ante sequential
equilibrium with the same distribution over outcomes. 
\end{thm}
\fullv{
\BPRF
Let $\vec{M}^1, \vec{M}^2, \ldots$ be a sequence of machine profiles
that converges to $\vec{M}$, and let $\mu$ be the belief induced by this
sequence. That is, $\mu$ is a belief that is compatible with $\vec{M}$. 
Let $\vec{M}$ be a NE that is not a sequential
equilibrium. 
There thus exists 
a player $i$ and 
a nonempty set of states $Q$ such that $M_i$ is not a best response for
$i$ at  
any state $q \in Q$, given the belief $\mu$. 
Let $q\in Q$ be a state that is not strictly preceded by another state 
$q^* \in Q$ (i.e., $q \notin \cQ_{M_i,q^*})$.
It
follows using the same proof as in Lemma \ref{lem:nashmixed0} that $q$
is reached with probability $0$ (when the profile $\vec{M}$ is used). 
Let $(M_i,q,M'_i)$ be a local variant of $M$ with 
the highest expected utility conditional on reaching $q$ and the other
players using 
$\vec{M}_{-i}$.  
(Since $\M$, the set of TMs, is finite, such a TM exists.)
Since $q$ is reached with
probability $0$, it follows that $((M_i,q,M'_i),\vec{M}_{-i})$ is a
NE; furthermore $M'_i$ is now optimal at $q$, and all states
that are reached with positive probability from $q$ (when using the
profile $((M_i,q,M'_i),\vec{M}_{-i})$ and belief system $\mu$).  

If $(M,q,M_i')$ is not a sequential equilibrium, 
we can iterate this procedure, keeping the belief system $\mu$
fixed.  Note that in the second iteration, we can choose only a state
$q'$ that is reached with probability 0 also starting from $q$ (when
using the 
profile $((M_i,q,M'_i),\vec{M}_{-i})$ and beliefs $\mu$). It follows by
a simple induction that states ``made optimal'' in iteration $i$ cannot
become non-optimal at later iterations. 
Since $\M$ is finite, it
suffices to iterate this procedure a finite number of time to
eventually obtain a strategy profile $\vec{M'}$ such that
$(\vec{M'},\mu)$ is a sequential equilibrium. 
\EPRF
}

We end this section by providing some existence results for games
with infinite machine spaces. 
As shown in Theorem \ref{nashimpliesseq.thm}, in games with positive
state cost, every NE is an ex ante sequential equilibrium. 
Although positive state cost is a reasonable requirement in many
settings, it is certainly a nontrivial requirement.
A game $G$ has {\em non-negative state cost} if the two
conditions in 
\fullv{Definition \ref{monotone.def}} 
\shortv{the definition of ``positive state cost''}
hold when replacing the
strict inequalities with non-strict inequalities. That is, roughly
speaking, $G$ has non-negative state cost if adding machine states
(without changing the functionality of the TM) can never improve the
utility. 
It is hard to imagine natural games with
negative state cost. 
In particular, a complexity function that assigns complexity 0 to all TMs
and inputs 
has non-negative state cost.
Say that $G$ is {\em complexity-independent} if, for each
player $i$, $i$'s utility does not depend on the complexity of players
$-i$.%
\footnote{For our theorem, it suffices to assume that player
$j$'s utility decreases if player $i$'s complexity decreases and
everything else remains the same.} 
(Note that all single-player games are trivially
complexity-independent.)
Although non-negative state cost 
combined with 
complexity-independence
is not enough to guarantee that every
NE is an ex ante sequential equilibrium (as illustrated by
Example~\ref{poscost.ex}), it is enough to guarantee the existence of 
an ex ante sequential equilibrium. 

\begin{proposition}
If $G$ is a 
complexity-independent
machine game with non-negative state
cost that has a NE, then it has a lean NE
with the same distribution over outcomes. 
\end{proposition}
\fullv{
\BPRF
Suppose that $\vec{M}$ is a NE of the game
$G$ with non-negative state cost. For each player $i$, let $M'_i$ denote
the TM obtained by removing all states from $M_i$ that are never
reached in equilibrium. 
Since $G$ has non-negative state cost
and is complexity-independent, $\vec{M'}$ is also a NE.
Furthermore, it is lean by definition and has the same 
distribution over outcomes as $\vec{M}$. 
\EPRF
}

\begin{corollary}
If $G$ is a 
complexity-independent machine
game with non-negative state cost, 
and $G$ has a NE,
then $G$ has an ex ante sequential equilibrium.  
\end{corollary}

\section{Extensive-form machine games}\label{sec:extensiveform}  
Up to now, we have 
considered sequential equilibrium only for Bayesian (machine) 
games, we have considered only
extensive-form games determined by Bayesian machine games, where players
make only one move in the underlying Bayesian game, and the remaining
moves correspond to computation steps.
But the notion of a machine game also can be extended to
extensive-form games as well in a straightforward way.  
We just sketch the relevant definitions here.

We assume that the reader is familiar with the standard definition of
extensive-form games.  We start with an underlying extensive-form game of
perfect recall.  The intuition here is that two nodes are in the
same information set for player $i$ if player $i$ cannot 
tree that the players could not distinguish the histories ending with
these nodes even if player $i$ recalled all the moves he has made made and all
the information that he has received.   Now if player $i$ chooses a TM that
forgets some information, then player $i$ may be able to make fewer
distinctions.  Thus, the information sets in the underlying game
represent an upper bound; player $i$'s actual partition may be coarser,
depending on his choice of TM.%
\footnote{The game in Figure~\ref{fig2} illustrates why we do not
want to start with a game of imperfect recall.  Recall that, in this
game, a player was able to use his strategy as a way of telling which
history in the information set he was is.  We want to separate the
fact that a player cannot distinguish two histories because the
information to distinguish is not available, even if he has perfect
recall, from the fact that he cannot distinguish two histories because
he has forgotten (more precisely, chosen to forget).  The former lack of
information is captured by the information sets of the underlying
extensive-form game; the latter is captured by the state of the TM.}

In an \emph{extensive-form machine game}, 
just as in the case of
computational Bayesian games, a player 
chooses a TM to play for him.  In addition to making moves in the
underlying game, the TM makes ``computational'' moves, just as in the
model of Section~\ref{sec:compex}.  In the resulting extensive-form game,
player $i$'s information sets are again determined by the states of the
TM that $i$ chooses.  Since all
we have are TMs, we need a way for a player's to make moves, to learn
about what moves other players made (if it is consistent with their
information sets to learn it), and to learn that it is their move.
We model this by assuming that players actually communicate with a 
mediator.  Formally, we use what are called \emph{Interactive} Turing
machines (ITMs), which can send and receive messages 
(see \cite{goldreich01} for
a formal definition.)  We assume that all communication passes between
the players and a trusted mediator.  Communication between the players
is modeled by having a trusted mediator who passes along messages
received from the players. Thus, we think of the players as having
reliable  communication channels to and from a mediator; no other
communication channels are assumed to exist.  The mediator is also an
ITM.  A player makes a move in the underlying game by sending the
mediator a message with that move; a player discovers it is his move and
gets information about other players' moves by receiving messages from
the mediator.  (See \cite[Section 3]{HP08} for details.)  With these
modifications, we can now define ex ante and interim
sequential equilibrium in the resulting extensive-form game just as in
Definition~\ref{seqeq.def}.   All our earlier results hold with
essentially no change.  
However, now an issue which did not seem to be so significant when
considering Bayesian games seems to have more bite when considering
extensive-form games.  TMs output bitstrings.  Thus, we have to
associate with bitstrings with actions in the underlying game.  Exactly
how we do this may affect the equilibrium.

\begin{example}\label{extensive.ex}
{\em Consider the (well-known) extensive-form game in Figure \ref{fig1}.}
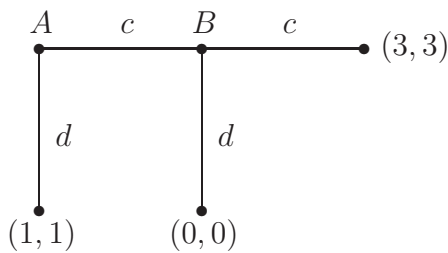
\begin{figure}[htb]
\begin{center}
\setlength{\unitlength}{.17in}
\begin{picture}(9,9)
\put(1,2){\circle*{.3}}
\put(6,2){\circle*{.3}}
\put(11,7){\circle*{.3}}
\put(1,7){\circle*{.3}}
\put(6,7){\circle*{.3}}
\put(1,7){\line(0,-1){5}}
\put(6,7){\line(0,-1){5}}
\put(1,7){\line(1,0){5}}
\put(6,7){\line(1,0){5}}
\put(3.5,7.5){$c$}
\put(8.5,7.5){$c$}
\put(1.5,4){$d$}
\put(6.5,4){$d$}
\put(0.7,7.5){$A$}
\put(5.7,7.5){$B$}
\put(0,1){$(1,1)$}
\put(5,1){$(0,0)$}
\put(11.5,6.8){$(3,3)$}
\end{picture}
\end{center}
\caption{A well-known extensive-form game.}
\label{fig1}
\end{figure}
{\em If we do not take computation into account, $(d,d)$ is a Nash
equilibrium, but it is not a 
sequential equilibrium: if Alice plays $c$, then Bob prefers switching
to $c$. The only sequential equilibrium is $(c,c)$
(together with the obvious belief assessment that assigns probability 1
to $(c,c)$).

To model the game in Figure \ref{fig1} as an extensive-form machine
game, we consider Alice and Bob communicating with a mediator
$\cN$. Alice sends its move to $\cN$; if the move is $c$, $\cN$ sends
the bit $1$ to Bob; otherwise $\cN$ sends $0$ to Bob. Finally, Bob sends
his move to $\cN$. Since the action space in a machine game is
$\{0,1\}^*$, we need to map bitstrings onto the actions $c$ and $d$. For
definiteness, we let the string $0$ be interpreted as the action $c$; all
other bitstrings (including the empty string) are interpreted as
$d$. Suppose that all TMs have complexity 0 on all inputs (i.e., computation
is free), and that utilities are defined as in Figure \ref{fig1}. 
Let $D$ be a 2-state machine that simply outputs $1$ (which is
interpreted as $d$); formally,
$D=(\tau,\{q_0,H\},q_0,\{H\})$, where $\tau$ is such that in $q_0$, the
machine outputs $1$ and transitions to $H$. 
Let $C$ be the analogous 2-state machine that outputs $0$ (i.e., $c$).} 
{\em
As we might expect, $((C,C),\mu)$ is both an interim and ex ante
sequential equilibrium, where 
$\mu$ is the belief assessment where 
Bob assigns probability 1 to receiving $0$ from $\cN$ (i.e., Alice
playing $c$). 
But now $((D,D),\mu')$ is also an interim and ex ante
sequential equilibrium, where $\mu'$ is the belief assessment where
Bob assigns probability 1 to receiving $1$ from $\cN$ (i.e., Alice
playing $d$). 
Since the machine $D$ never reads the input from $\cN$, 
this belief is never contradicted.  We do not have to consider what his
beliefs would be if Alice had played $c$, because he will not be in 
a local state where he discovers this.
$(C,C)$ remains a sequential equilibrium even if we charge (moderately)
for the number of states.  
$(D,D)$, on the other hand, is no longer a sequential equilibrium, or
even a Nash equilibrium: both players prefer to use the single-state
machine $\bot$ that simply halts (recall that outputting the empty
string is interpreted as choosing the action $d$). Indeed, 
$(\bot,\bot)$ is an ex ante and interim sequential equilibrium. 
}
\qed
\end{example}
As illustrated in Example~\ref{extensive.ex}, to rely on our treatment
of sequential equilibrium, we must first interpret an extensive form
game as a mediated game. 
But as we see, the sequential equilibrium outcomes are sensitive
to the interpretation of the extensive-form game. This leaves open the
question of what the ``right'' way to interpret a given extensive-form
game is.

\section{Discussion}\label{sec:discussion}

We have given definitions of ex ante and interim sequential equilibrium
in machine games, provided conditions under which they exist, and
related them to Nash equilibrium in machine games.  We believe that
thinking about sequential equilibrium in machine games clarifies some
issues raised by Piccione and Rubinstein \citeyear{PR97} about
sequential equilibrium in games of imperfect recall. 

Specifically, given a ``standard'' extensive-form game $G$ of imperfect
recall, we can consider an extensive-form computational game $G'$ where
the players have the same moves available as in $G$, but $G'$ is a game
of perfect recall.  We can then
restrict 
the class of TMs that the agents can choose among to those
to the computable convex closure\footnote{See \cite{HP08} for a formal
  definition of the computable convex closure.} of a finite set of TM
that capture the knowledge assumptions described by the information sets
in $G$.    In particular, if nodes $x$ and $y$ are in the same
information set for player $i$ in $G$, then all the TMs that $i$ can
choose among in $G'$ will be in the same state when they reach both
nodes $x$ and $y$.  
We restrict to complexity functions where
randomization is free, in the sense that the complexity of $\alpha M_1 +
(1-\alpha)M_2$ is the obvious convex combination of the complexity of
$M_1$ and the complexity of $M_2$.  As shown in \cite{HP08}, this
suffices to guarantee that $G'$ has a Nash equilibrium.  By
Theorem~\ref{nashimpliesseq.thm}, if $G'$ has positive state cost, then
$G'$ has an ex ante sequential equilibrium; furthermore, if the
complexity function in $G'$ is local, by Proposition~\ref{pro:obvious},
this ex ante sequential equilibrium is also an interim sequential
equilibrium.  Thus, thinking in terms of computational games forces us
to specify the ``meaning'' of the information set a game of imperfect
recall, and gives us a way of doing so.  It also gives us deeper insight
into when and why sequential equilibrium exists in such games.

Thinking in terms of computational games also raises a number of
fundamental questions involving 
computation in extensive-form games.   As we already observed in
\cite{HP08}, in our framework, we are implicitly assuming that the
agents understand the costs associated with each TM; they do not have to
compute these costs.  Similarly, players do not have to compute their
beliefs.  In a computational model, it seems that we should be able to
charge for these computations.  It is not yet clear how to charge for
these computations, nor how such charges should affect solution
concepts.  We are planning to explore these issues.

\bibliography{z,joe}
\end{document}